\documentclass[aip,cha,amsmath,amssymb,reprint]{revtex4-1}

\usepackage{graphicx}
\usepackage{dcolumn}
\usepackage{bm}
\usepackage[utf8]{inputenc}
\usepackage[T1]{fontenc}
\usepackage{mathptmx}
\usepackage{color}

\newcommand{\rem}[1]{}



\begin{document}
\title{Quasi-stationary states of game-driven systems: a dynamical approach} 
\author{Sergey Denisov}
\email[Corresponding author: ]{sergiyde@oslomet.no}
\affiliation{Department of Computer Science, Oslo Metropolitan University, N-0130, Oslo, Norway}
\author{Olga Vershinina}
\affiliation{Department of
Applied Mathematics, Lobachevsky University, 603950, Nizhny
Novgorod, Russia}
\author{Juzar Thingna}
\affiliation{Center for Theoretical Physics of Complex Systems (IBS),  Daejeon 34126, South Korea}
\author{Peter H\"{a}nggi}
\affiliation{Institut f\"ur Physik, Universit\"at Augsburg,
D-86135 Augsburg, Germany}
\author{Mikhail Ivanchenko}
\affiliation{Department of
Applied Mathematics, Lobachevsky University, 603950, Nizhny
Novgorod, Russia}
\date{\today}
\begin{abstract}
Evolutionary game theory is a framework to formalize  the
evolution of collectives (``populations'') of competing agents that
are playing a game and,  after every round, update their strategies to maximize individual payoffs. 
There are two complementary approaches to modeling evolution of player populations.  The first 
addresses essentially finite populations by implementing the apparatus 
of Markov chains. The second assumes that the populations are infinite  and operates with a system of
mean-field deterministic differential equations.  By using a model of two antagonistic  populations, which are
playing a game with stationary or periodically varying payoffs, we demonstrate that it
exhibits metastable dynamics that is reducible neither to an immediate transition to a fixation (extinction of all but one strategy in a finite-size population) 
nor to the  mean-field picture. In the case of stationary payoffs, this dynamics can be  captured with 
a system of stochastic differential equations  and interpreted as a stochastic Hopf bifurcation.
In the case of varying payoffs, the metastable dynamics is much more complex than the dynamics of the means.
\end{abstract}
\maketitle
\begin{quotation}
Although first formalized by Darroch and Seneta in 1964 \cite{darroch},
the best (in our opinion) outline of quasi-stationary states (QSs) was given by 
Yaglom when he asked in 1947 \cite{yaglom}: ``\textit{We are dealing with a stochastic process with an absorbing state (i.e., a state which cannot
be left once the system got into it) so that the absorption happens with probability one. What is
the distribution in the limit $t\rightarrow \infty$ provided that the absorption did not happen until time $t$?}''.
In the Markov chain framework, the QSs are related to the probability distributions over the set of states from which all the absorbing \cite{madsen} ones 
are excluded. Some quasi-stationary distributions could be very sustainable so that the time needed for their eventual 'evaporation' into the absorbing states 
is much longer than all the relevant observation time scales~\cite{dorn}. In this case the QS is also a \textit{metastable} state \cite{meta1}.
The QS concept naturally applies to the game-driven evolution of finite populations \cite{traulsen1,traulsen3}. Quasi-stationary states were recently interpreted as a mean \cite{ge} 
to resolve a conflict \cite{traulsen1} between the stochastic approach  based on the Markov chain ideology \cite{traulsen3}  and 
deterministic mean-field description \cite{hofbauer}. Namely, while the former tells
that the asymptotic regime of any finite  population is a fixation, the latter yields an asymptotic solution  either in the form of a periodic 
cycle or a chaotic attractor or a fixed point (an 'evolutionary stable strategy' in which different types of players are represented \cite{hofbauer}). 
The gap between the two mutually exclusive pictures could be bridged with QSs, by observing that, in the case of large but finite populations, 
the corresponding distributions are  localized around the mean-field solutions and declaring that the mean-field solution is thus transformed 
into a transient (metastable) dynamics~\cite{ge,ge1,herpich18}. 
Here we demonstrate that this is not always the case and  the metastable dynamics, underlying the QSs, can be  
very different from the mean-field solution.
\end{quotation}

\section{Introduction} 
The agenda of Evolutionary  Game  Theory (EGT) is to explain the evolution of biological populations by formalizing two main driving factors, 
conflict and cooperation~\cite{maynard}. Current applications of the theory range far beyond the original biological context
and it is used to model dynamics of financial markets~\cite{market} and interpret condensation phenomena in dissipative quantum systems~\cite{frey1}.

The central element of EGT is the interaction of players during a round of a game. 
Although formally any number of players can be involved into the interaction \cite{multiverse}, 
most of the existing results were obtained for two-player games. Players can choose strategies from  fixed sets 
and their interaction is mediated by payoffs corresponding to the choices they made. For a two-player game with  two strategies 
per player, the payoffs can be  arranged in a bi-matrix \cite{hofbauer}, 
\begin{eqnarray} 
\begin{bmatrix}
       a_{11},b_{11} & a_{12},b_{21} \\
       a_{21},b_{21} & a_{22},b_{22}      
     \end{bmatrix}.
\label{Eq:bimatrix}
\end{eqnarray}
For example, if player $A$ chooses strategy $1$ and player $B$ chooses strategy $2$, the former receives payoff $a_{12}$ and the latter
receives payoff $b_{21}$. After every round,  players monitor the payoffs obtained by their peers (other members of the population) 
and try to adapt strategy of the most successful ones.
There are several formalizations of the strategy adaptation process; the Moran process \cite{moran,moranNature} is currently the 
most popular one; see, e.g., Refs.~[\onlinecite{traulsen1,ge,ge1,garcia}].

Finite sizes of animal  populations favor stochastic Markovian approaches~\cite{madsen,nowak} to modeling game-driven evolution. By assuming that players 
belonging to the same population are indistinguishable, the iterative process of matching and consequent strategy adaption 
can be recast in the form of a Markov chain. 
The state of a population is then fully specified by the  probability vector which assigns probability to every possible arrangement of the strategies.
In the absence of mutations~\cite{mutations} (the case we address here), a state corresponding to the situation when the whole population 
uses the same strategy, is an \textit{absorbing state}~\cite{markov}. 
Once the population enters this state, a fixation is happened~\cite{nowak}.

An alternative approach, 
which was prevailing in the EGT field until recently, addresses the case of  infinite populations. 
Its main tool is the celebrated replicator model, which is a system of nonlinear deterministic differential equations~\cite{hofbauer}.
The variables governed by these equations are relative frequencies of the strategies  which are assumed to be continuous probabilities.
Because of the non-linearity of its equations, the replicator model is able to exhibit a spectrum of dynamical regimes, 
ranging from fixed point solutions to periodic oscillations and chaos (when the number of players and/or number strategies per player 
are larger than two; see, e.g., Refs.~[\onlinecite{chaos1,chaos2,chaos3}]).

In between these two limits there is a land of large but finite populations. Many animal populations fall into 
this category and so analysis of the corresponding models 
can provide some additional  insight into complex real-life phenomena~\cite{traulsen33}. It is intuitive that finite size fluctuations play an important 
role in the evolutionary dynamics of such populations but these populations are too big to be modeled with Markov chains~\cite{note}. 
The diffusion approximation of Markov processes~\cite{gardiner}, is an appealing tool to bridge the two approaches and explore the land between them. 
When implemented to Moran processes \cite{moran,moranNature}, the approximation yields a system of  stochastic differential equations (SDEs) 
with a multiplicative cross-correlated noise~\cite{traulsen1,traulsen44}. The noise strength scales down as the population size 
increases so that, in the thermodynamic limit, the equations reduce to the deterministic replicator model. 

Because the size of the population enters the SDEs as a parameter, the diffusion approach provides a tremendous speed-up
as compared to Markov chain simulations and allows for simulating models of an arbitrary large size~\cite{traulsen44}.
The SDE approach can be used to resolve QSs and capture the transient (before the fixation) dynamics, e.g., by replacing the absorbing boundary conditions 
with the reflecting ones~\cite{ge1} and then performing a routine ensemble averaging.

Here we apply the concept of quasi-stationary distributions (QDs) ~\cite{darroch,QS} to analyze evolutionary dynamics of two antagonistic populations 
driven by a  two-player game. We consider two variants of the game, with stationary payoffs
and with payoffs changing from round to round, in a periodic manner (this choice is motivated by some biological phenomena; see the next section).

We demonstrate that, in both cases, the metastable dynamics underlying the QDs is very different from the 
solutions of the corresponding replicator models. In the stationary case, the latter is a fixed point while 
the metastable dynamics is manifested by a stochastic limit cycle, which, within the SDE framework, can be interpreted as a result of 
a stochastic Hopf bifurcation \cite{arnold,baxendale,kurths}. 
By employing the Floquet theory~\cite{floquet,floquetH}, we generalize the notion of QD to evolutionary games with periodically varying payoffs. 
We demonstrate that the corresponding non-equilibrium states  are result of a complex dynamics which cannot be reduced to the evolution of means. 

\section{Model} 

\begin{figure}[t]
\includegraphics[width=0.45\textwidth]{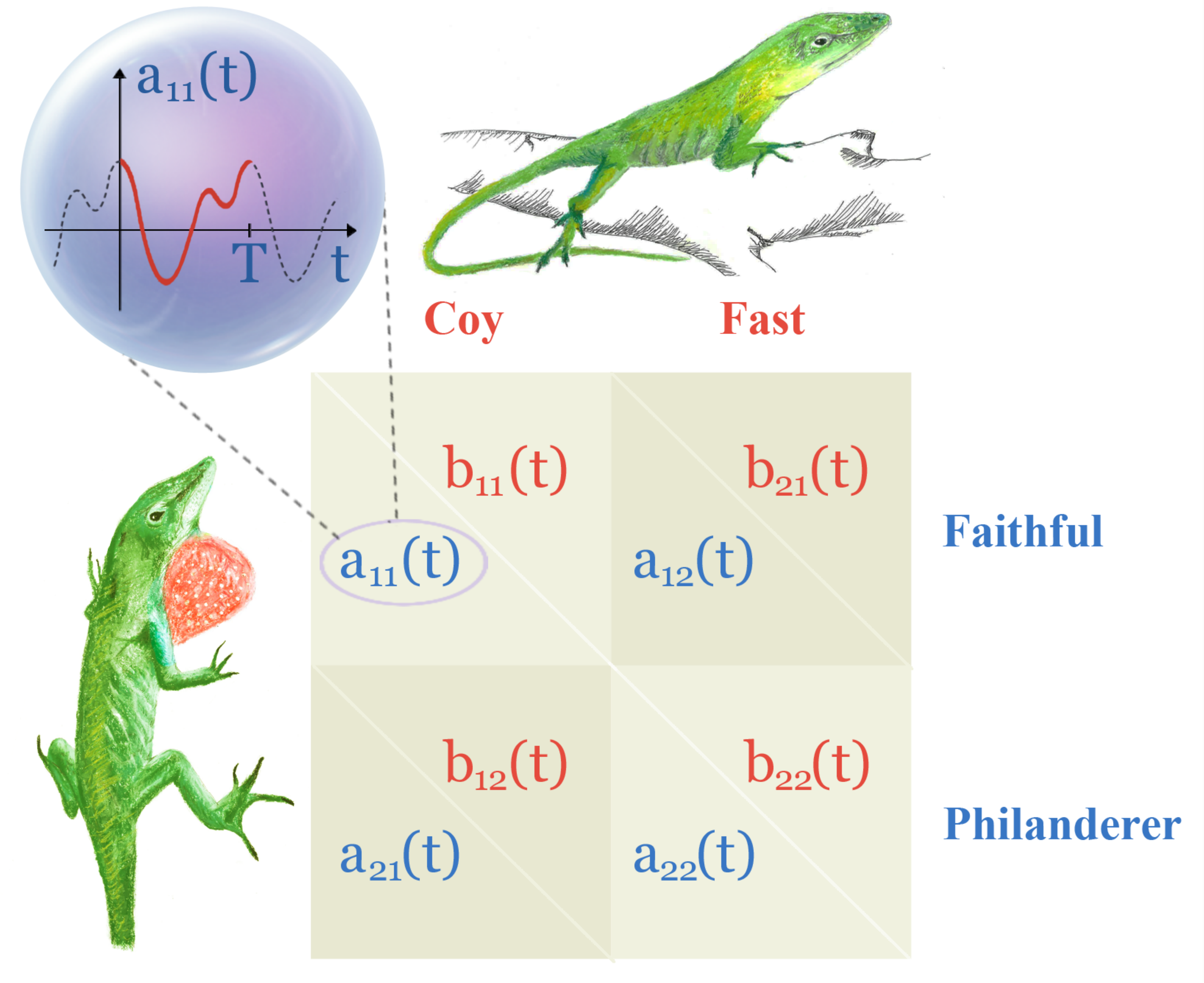}
\caption{(color online) \textbf{`Battle of Sexes' with seasonal variations of mate preferences. }
It can model, e.g.,  the process of mate selection in a population of Carolina anole  lizards. 
A female of the species could be either \textit{coy} (and prefer a long and arduous courtship, in order to be sure 
that a  mate will contribute to parental care) or \textit{fast} (and not worry much about parental care). A male could 
be either \textit{faithful} (and ready to assure a potential partner that he will be a `faithful husband' by performing 
a long courtship) or \textit{philanderer} (and thus would prefer to shorten the courtship stage). Depending on the strategies, $s$ (played by the female) 
and $s'$ (played by her mate), the female gets payoff $b_{ss'}$ (red) and the male gets $a_{s's}$ (blue). Both, females and males, are 
season-constrained in their preferences of opposite-sex mates. These seasonal constraints are modeled via  time-periodic variations of the payoffs.
}\label{Fig:1}
\end{figure}

Motivated by the original biological context of EGT, we decided to use the celebrated ``Battle of Sexes'' model by Dawkins \cite{dawkins}, 
as an example. This model formalizes a sex conflict over the parental investment \cite{sexual} and two antagonistic populations 
are represented by females and males of a species  (see sketch in Fig.~\ref{Fig:1}).
Players of each sex have two strategies and payoffs are given by a $2 \times 2$  bi-matrix. We generalize this model
by introducing periodic time variations in the payoffs. We are not only inspired by the possibility to encounter a more complex evolutionary dynamics
but also by several new findings in ecology.

In recent years it has been found that in many species mating strategies and preferences are not constant in time but are 
season-dependent \cite{fly,crab,goby,molly}. When courting (selecting) a mate, a female or male of the species faces a complex choice problem 
when benefits of a choice depend on the season and have to be traded off against each other in the context of current environmental conditions. 
For example,  the rate of the hormonal activity, 
courtship strategies, and mate selection of Carolina anole lizards (\textit{Anolis carolinensis}), both of females and males, 
are regulated by the temperature and photo-period and thus are strongly season dependent \cite{lizard1}. 
Even the amount of different types of muscle fibers that control the vibrations of a red throat fan (dewlap) - which males employ during courtship - 
is season dependent \cite{lizard2}. Within the ``Battle of Sexes'' framework, this can be modeled by introducing periodically varying payoffs,  
as illustrated in Fig.~\ref{Fig:1}.

Namely, players $A$ (males) and $B$ (females) form two populations of a fixed size $N$, with each of them having two strategies $s = \{1,2\}$. 
Payoffs, $\{a_{ss'}\}$ and $\{b_{s's}\}$, $s,s' = \{1,2\}$, may change from round to round.
During a single round, members of the antagonistic populations
are randomly matched and then simultaneously play $N$ games. After that, a strategy adaptation phase takes place in every populations.
Then the process is iterated.

This evolution is an essentially  discrete-time process and its rounds are labeled with index $m$. 
In order to be able to compare the discrete-time evolution to the mean-field dynamics,
we introduce time variable $t$, which is counted from zero and incremented by $\Delta t$ after every round.
Now we define time-periodic payoffs, $c_{ss'}(t) = c_{ss'}(t+T)$, $c = \{a,b\}$, where  $\Delta t = T/M$.
The payoffs can be represented as sums of constant and zero-mean time-periodic  components, 
$c_{ss'}(t) = \bar{c}_{ss'} + \tilde{c}_{ss'}(t)$, $\langle\tilde{c}_{ss'}(t)\rangle_T = 0$. 
After $M$ rounds the payoffs return to their initial values. 

To introduce the strategy adaptation stage, we use 
the Moran process \cite{moran, moranNature}, which works in the following way.
The state of the populations after the $m$-th round is specified by 
the number of players playing first strategy from their strategy list, $i$ (males with $s=1$) and $j$ (females with $s'=1$), $0 \leq i,j \leq N$. 
The payoffs obtained by the players using strategy $s$ are
\begin{eqnarray}
\pi^{A}_s(j,t) &= a_{s1}(t)\frac{j}{N}+a_{s2}(t)\frac{(N-j)}{N}, \\
\pi^{B}_s(i,t) &= b_{s1}(t)\frac{i}{N}+b_{s2}(t)\frac{(N-i)}{N}.
\label{Eq:quasi_sym1}
\end{eqnarray}
Payoffs determine the probabilities for a player to be chosen for reproduction, e.g., for population $A$, 
\begin{eqnarray}
P^{A}_s(i,j,t) = \frac{1}{N} \cdot \frac{1 - w + w\pi^{A}_s(j,t)}{1 -w + w \bar{\pi}^{A}(i,j,t)},
\label{Eq:quasi_sym2}
\end{eqnarray}
where $\bar{\pi}^{A}(i,j,t) = [i\pi^{A}_{1}(j,t)+(N-i)\pi^{A}_{2}(j,t)]/N$ is the average payoff for the population $A$. 

The baseline fitness $w \in [0,1]$ is a tunable parameter of the game driven evolution~\cite{moranNature,traulsen1}. 
For example, when $w = 0$, the probability to be chosen for reproduction does not depend on player's payoff and is 
uniform across the population. After the choice for reproduction has been made, another member of the population is chosen 
completely randomly and replaced with an offspring of the chosen player, i.e. with a player 
which uses the same strategy as its parent \cite{imitation}. This  mechanism is acting simultaneously in 
both populations, $A$ and $B$, so that one male and one female is chosen 
for reproduction. The mating pair then produces two offspring, a male and female, which then introduced into the corresponding populations. 
The size $N$ of both population is therefore preserved.

A single round can be considered as a two-dimensional Markov chain, a generalization of the one-dimensional chain introduced in Ref.~\cite{traulsen1, traulsen44}. Before formalising the Markov chain,
we first define the transition rates for two populations of players.
For example, the probability for population $A$ to get one more player with strategy $1$ after one round  (and, correspondingly, one player with strategy $2$ less) is given by
\begin{eqnarray} 
T_{A}^{+}(i,j,t) =\frac{1-w + w\pi^{A}_1(t)}{1 -w + w \bar{\pi}^{A}}\frac{i}{N}\frac{N-i}{N}, 
\label{Eq:rates1}
\end{eqnarray}

The probability  to get one more player with strategy $2$   (and, correspondingly, one player with strategy $1$ less) is
\begin{eqnarray}
T_{A}^{-}(i,j,t) =\frac{1-w + w\pi^{A}_2(t)}{1 -w + w \bar{\pi}^{A}}\frac{N-i}{N}\frac{i}{N}.
\label{Eq:rates2}
\end{eqnarray}
The probability for population $B$ are defined in the similar manner. These 
four probabilities are building blocks to construct the transition matrix for the two-dimensional Markov process (see Appendix).

By following the procedure  from Ref.~[\onlinecite{traulsen1}], it can be shown that, in the limit $N \rightarrow \infty$ and $dt = \frac{1}{N} \rightarrow 0$, 
the dynamics of the variables $x = i/N$ and $y = j/N$ is governed by the adjusted replicator equations \cite{maynard,hof2},
\begin{align}
\label{Eq:replicatorx}
\dot{x} &= x[1-x][\vartriangle^A(t) + \Sigma^A(t)y]\frac{1}{\Gamma + \bar{\pi}^A(x,y,t)}, \\
\label{Eq:replicatory}
\dot{y} &= y[1-y][\vartriangle^B(t) + \Sigma^B(t)x]\frac{1}{\Gamma + \bar{\pi}^B(x,y,t)},
\end{align}
where $\vartriangle^{C} = c_{12} - c_{22}$, $\Sigma^{C}=c_{11}+c_{22}-c_{12}-c_{21}$, $\Gamma = \frac{1-w}{w}$, and $C(c)=\{A(a), B(b)\}$. 

Within the stochastic framework, a single round can be represented as the multiplication \cite{transition} 
of the state $\mathbf{p}$, expressed as a $N \times N$ matrix with elements $p(i,j)$, 
with the transition fourth-order tensor $\mathbf{S}$, with elements $S(i,j,i',j')$ (see Appendix). 
By using the bijection $k = (N - 1)j + i$, we can unfold the probability matrix $p(i,j)$ 
into the vector $\tilde{p}(k)$, $k = 0,...,N^2$, and the tensor $S(i,j,i',j')$ into the matrix $\tilde{S}(k,l)$ with $k,l = 0,1,...,N^2$. 
This reduces the problem to a Markov chain \cite{gant}, $\tilde{\textbf{p}}^{m+1} = \tilde{\textbf{S}}^{m}\tilde{\textbf{p}}^{m}$, where $m$ is 
the round to be played and every state is fully specified now with a single integer $k$. 

The four states $(i = \{0,N\}, j= \{0,N\})$ are absorbing states because the transition probabilities leading out of them, Eqs.~(\ref{Eq:rates1}-\ref{Eq:rates2}), are identically zero. The absorbing states are therefore attractors (sinks) of the evolutionary dynamics, and it is evident that  finite-size fluctuations will eventually drive a population to one of these states. 
\begin{figure*}[t]
\includegraphics[width=0.95\textwidth]{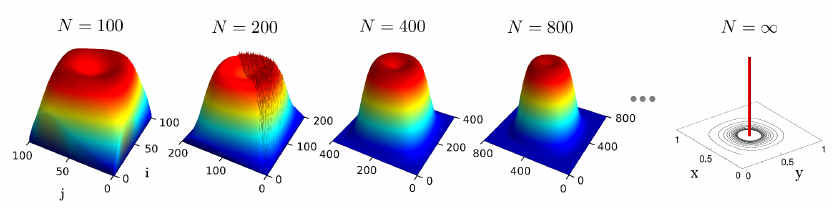}
\caption{(color online) \textbf{Quasi-stationary probability distributions of the `Battle of Sexes'  with constant payoffs.}
For all values of $N$, the distribution were obtained by finding the maximal-eigenvalue eigenvector of the reduced 
transient matrix $\tilde{\mathbf{Q}}^{1}$, Eq.~(\ref{Eq:supermatrix}) \cite{arpack}.
For $N=200$ the plot combines the distribution obtained by reshuffling the eigenvector  (left half) and the histogram sampled  by using 
Langevin equations (\ref{Eq:Langevinx} - \ref{Eq:Langeviny}), which were propagated to time $t_{fin} = 10^6$. (right half).  
The total number of realization is $10^7$. 
In the thermodynamic limit $N = \infty$ (right panel), 
the quasi-stationary distribution shrinks to the Nash equilibrium $\left(\frac{1}{2},\frac{1}{2}\right)$, which thus  becomes 
a fixed point of the mean-field dynamics. 
The baseline fitness $w = 0.3$. 
}\label{Fig:2}
\end{figure*}

We are interested in the dynamics before the absorption, so we merge the four states into a \textit{single} 
absorbing state by summing the corresponding incoming rates. The boundary states, $(i = \{0,N\},j\in \{1,\cdots,N-1\})$ and $(i\in \{1,\cdots,N-1\},j=\{0,N\})$, 
can also be merged into this absorbing `super-state'. The reason for including these additional states into the `absorbing' state is that once the population gets to the  boundary, it will 
only move towards one of the two nearest absorbing states. By labeling the absorbing super-state with index $k = 0$, 
we end up with a stochastic $(L+1) \times (L+1)$ matrix
\begin{eqnarray} 
\tilde{\mathbf{S}}^{m} = \begin{bmatrix}
       1 & \boldsymbol{\varrho_0}^{m} \\
       \mathbf{0} & \tilde{\mathbf{Q}}^{m},      
     \end{bmatrix}
\label{Eq:supermatrix}
\end{eqnarray}
where $L = (N-2)^2$, $\mathbf{\varrho}_0^{m}$ is row vector given by the set of 
incoming transition probabilities to the absorbing super-state, $\mathbf{0}$ is 
a $L\times 1$ column vector will all entries zero, and $\tilde{\mathbf{Q}}^{m}$ is a $L \times L$ reduced transition matrix.

With Eq.~(\ref{Eq:supermatrix}), we arrive at the the setup used by Darroch and Seneta to 
formulate the concept of QDs \cite{darroch}. 
Namely, there exists a vector $d$, which represent a quasi-stationary distribution (QD). 
This distribution can be very sustainable and remains near invariant under the action of the matrix $\tilde{\mathbf{S}}^{m}$. 
In this case, it is a metastable state \cite{meta1}. The QD can be obtained from 
the  right eigenvector $d$ of the reduced transition matrix $\tilde{\mathbf{Q}}^{m}$ 
corresponding to the maximum (by modulus) eigenvalue $\lambda$ \cite{darroch}.  By virtue of the Perron-Frobenius theorem, $\lambda$ is 
real and vector $d$ is real and non-negative \cite{gant}. 
By applying the inverse bijection  and then performing the normalization,  we obtain two-dimensional probability 
distribution~$\mathbf{d}(x,y)$.

The most straightforward way to evaluate of the dynamics underlying the QDs is to perform  Monte-Carlo sampling. 
To address  the limit $N \gg 1$, we follow the recipe given in Refs.~[\onlinecite{traulsen1,traulsen44}] and 
derive stochastic differential equations (SDEs), which can be then used to approximate the evolution.
As the first step, we introduce continuous variables $x = i/N$, $y = j/N$, and $t = m/N$. Next we define probability density $\varrho(x,y,t) \backsimeq  
N\cdot p_m(i, j)$. Finally, by Taylor expanding the Markov chain in orders of $1/N$ and truncating the expansion 
after the  second order, we obtain a Fokker-Planck equation \cite{risken}
\begin{widetext}
\begin{align}
\dot{\rho}(x,y,t) & = -\sum_{k=A,B} \frac{\partial}{\partial x_{k}}\cdot\left[\nu_{k}(x,y,t)\rho(x,y,t)\right]+\frac{1}{2}\sum_{k,l=A,B}\frac{\partial^{2}}{\partial x_{k}x_{l}}\cdot \left[d_{kl}(x,y,t)\rho(x,y,t)\right],
\label{Eq:FP}
\end{align}
\end{widetext}
where
\begin{eqnarray}
\nu_{k}(x,y,t) &=& T^{+}_{k}(x,y,t) - T^{-}_{k}(x,y,t),\\
d_{kk}(x,y,t) &=& \frac{T^{+}_{k}(x,y,t) + T^{-}_{k}(x,y,t)}{N},\\
d_{kl}(x,y,t) &=& \frac{\nu_{k}(x,y,t) \cdot \nu_{l}(x,y,t)}{N}, \quad \quad k \neq l,
\label{Eq:coefLangevinEqs}
\end{eqnarray}
where $k,l \in \{A, B\}$.
The corresponding stochastic differential equations can be represented in
the Langevin form \cite{risken} 
\begin{eqnarray}
\label{Eq:Langevinx}
\dot x(t) = \nu_{A}(x, y, t) + \sum\limits_{l = A,B}g_{Al} \cdot \xi_{l}(t),\\
\label{Eq:Langeviny}
\dot y(t) = \nu_{B}(x, y, t) + \sum\limits_{l = A,B}g_{Bl} \cdot \xi_{l}(t),
\end{eqnarray}
where $\xi(t)$ is an uncorrelated Gaussian white noise of variance one. 
The $2 \times 2$ matrix 
$G = \begin{pmatrix}
	g_{AA} & g_{AB} \\
	g_{BA} & g_{BB} \\
\end{pmatrix}
$ can be expressed in terms of the diffusion matrix
$D = \begin{pmatrix}
	d_{AA} & d_{AB} \\
	d_{BA} & d_{BB} \\
\end{pmatrix}
$ via $G^TG = D$.
We integrate equations (\ref{Eq:Langevinx} - \ref{Eq:Langeviny}) by using the standard Euler - Maruyama method \cite{euler} with time step $dt = 10^{-4}$.
In order to obtain matrix $G$, on every step we diagonalize $2 \times 2$ diffusion 
matrix $D = U \Lambda U^{T}$ and use its eigenvectors to construct matrix $G = U \sqrt{\Lambda} U^{T}$. A conditional sampling of QD 
is performed by integrating the Langevin equations up to time $t_{fin}$ and updating the histogram with the final points of the trajectories.

\section{Stationary payoffs: a stochastic bifurcation} 

We first consider the case  when all payoffs are constant. 
As an example, we use a game with  payoffs $a_{11}$, $a_{22}$, $b_{12}$ and $b_{21}$ equal $1$, 
and $-1$ for the rest of strategies (this choice corresponds to the  `Matching Pennies' game~\cite{neumann}). 
Figure~\ref{Fig:2} presents the metastable states of the corresponding evolution. 
In order to find them we numerically obtain maximal-eigenvalue eigenvector  of the matrix $\tilde{\mathbf{Q}}^{1}$ ($m=1$ since 
the payoffs are constant) and perform sampling by using the SDEs,~Eqs.~(\ref{Eq:Langevinx},\ref{Eq:Langeviny}). 

\begin{figure}[t]
\includegraphics[width=0.48\textwidth]{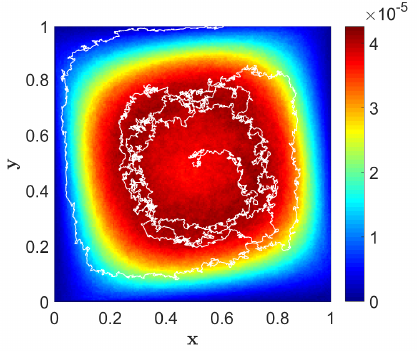}
\caption{\textbf{Transient evolution of two populations driven by `Battle of Sexes' game with constant payoffs.}
The color indicates the quasi-stationary distribution $\mathbf{d}(x,y)$. 
The white line corresponds to a single realization of the Markov process initiated at the point $(\frac{N}{2},\frac{N}{2})$.
$N = 200$. Other parameters are as in Fig.~\ref{Fig:2}.}\label{Fig:3}
\end{figure}

The means (first moments of the QD),
\begin{eqnarray}
\bar{x} = \sum_{i,j=1}^{N-1} \frac{i\cdot d(i,j)}{(N-2)^2};~\bar{y} = \sum_{i,j=1}^{N-1} \frac{j\cdot d(i,j)}{(N-2)^2},
\label{Eq:means}
\end{eqnarray}
coincide with the Nash equilibrium $\left(\frac{1}{2},\frac{1}{2}\right)$ for any $N$. 
Yet the distributions obtained for $N=100$ and $N=200$ are not simply peaked but  have `craters' on their tops. 
Stochastic Markovian simulations reveal the existence of a long transient trajectory going along the ridge of the craters and encircling the Nash 
equilibrium; see~Fig.~\ref{Fig:3}. The SDE integration yields a near identical dynamics.
However, the SDE framework offers an interpretation of the dynamics as a limit cycle resulted from a stochastic Hopf 
bifurcation~\cite{arnold,baxendale,kurths,hopf1}. 
The latter is a result of a marginal stability of the fixed point $(\frac{1}{2},\frac{1}{2})$ of 
the replicator system~(\ref{Eq:replicatorx} - \ref{Eq:replicatory})~\cite{unpublished}.

\begin{figure}[t]
\includegraphics[width=0.48\textwidth]{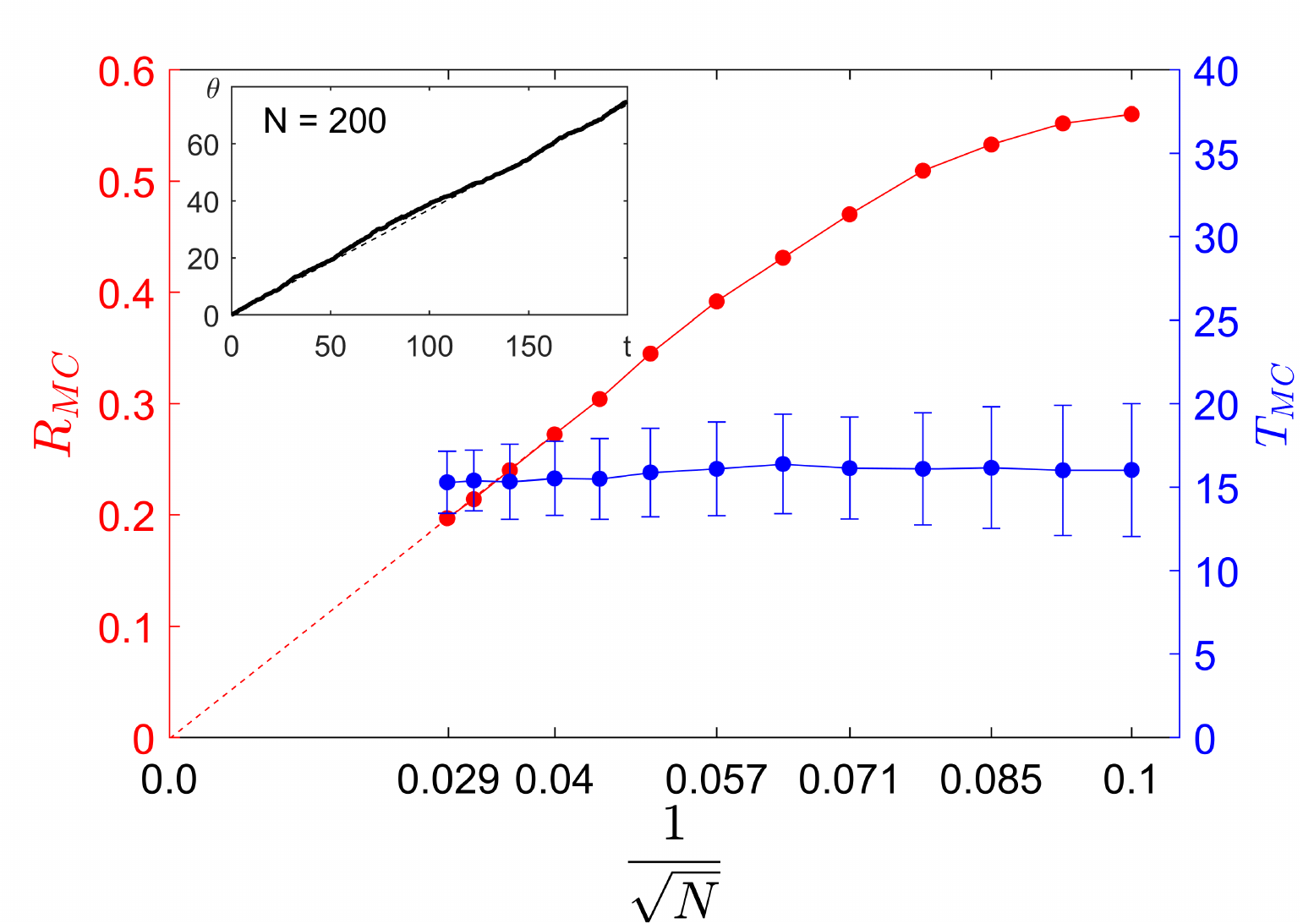}
\caption{\textbf{Radius of the metastable limit cycle (red) and its period (blue) for the `Battle of Sexes' 
with constant payoffs.} For  $N \leq 300$ ($1/\sqrt{N}\geq 0.057$) both quantities were obtained from  quasi-stationary distributions and Monte-Carlo sampling. 
For $N > 300$, a sampling with the SDEs, Eqs.~(\ref{Eq:Langevinx} - \ref{Eq:Langeviny}), was used. 
Inset shows the evolution of the  polar angle (black line) of the stochastic trajectory of the  
stochastic differential equations (\ref{Eq:Langevinx},\ref{Eq:Langeviny}). Black dashed line corresponds to linear dependence $\langle \Omega\rangle t$, 
where $\langle\Omega\rangle=2\pi/T_{MC}$ for $N=200$. 
Red dashed line indicates an inverse dependence of the cycle radius w.r.t. the fluctuation strength $\sqrt{N}$. 
The radius scaling and the period independence of $N$ are signatures of a stochastic Hopf bifurcation \cite{hopf1}. 
All parameters are the same as in Fig.~\ref{Fig:3}.}
\label{Fig:4}
\end{figure}

\begin{figure}[b]
\includegraphics[width=0.4\textwidth]{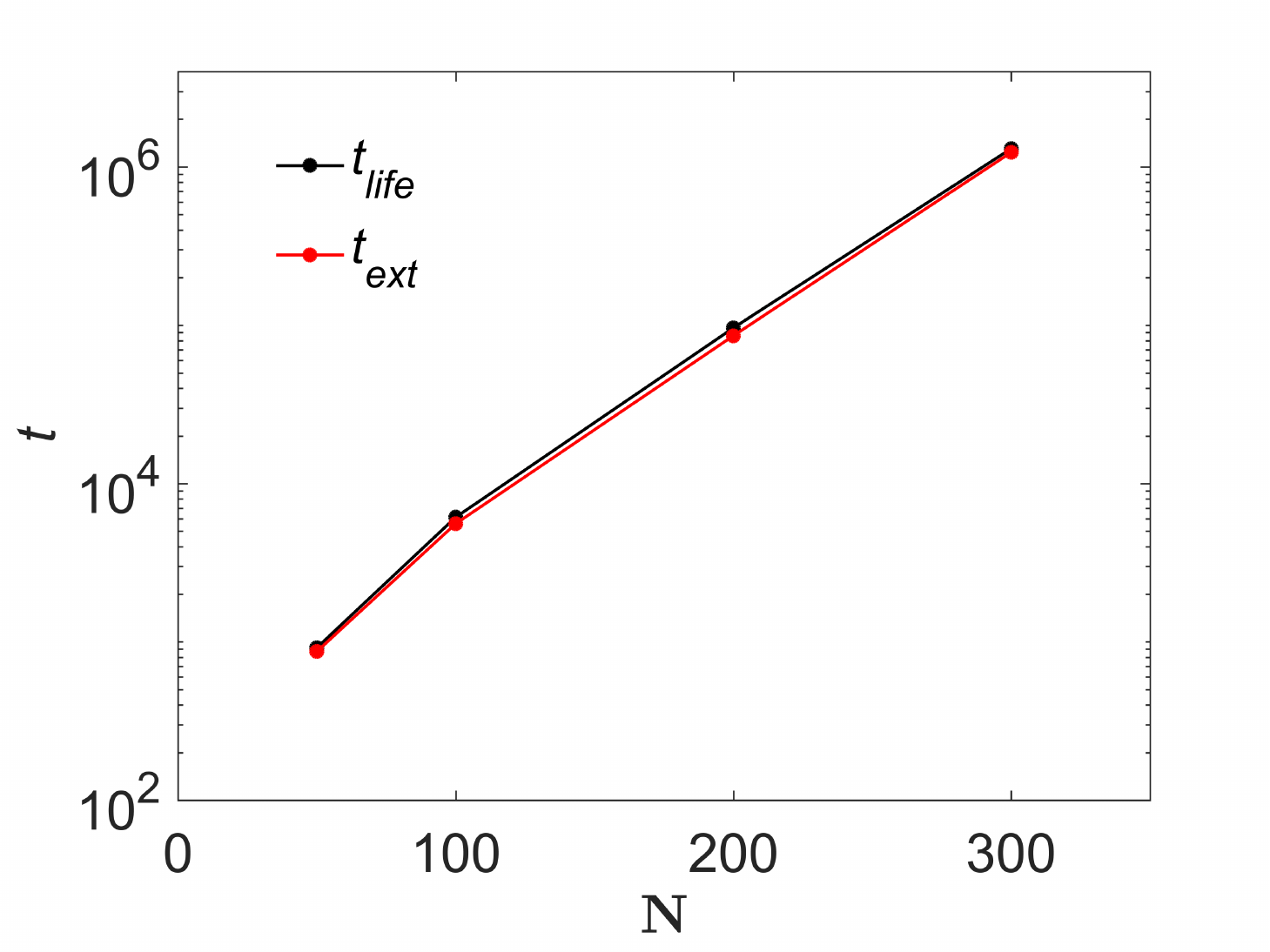}
\caption{\textbf{The average lifetime $t_{\mathrm{life}}$ vs average extinction time $t_{\mathrm{ext}}$ for the `Battle of Sexes' 
with constant payoffs.} The average extinction time was sampled with $10^5$ realizations per non-absorbing state.
All parameters are the same as in Fig.~\ref{Fig:3}.
}\label{Fig:5}
\end{figure}

The limit cycle character of the dynamics can be validated by tracking the trajectory polar angle as a function of time and 
noting a well pronounced linear dependence; see inset on Fig.~\ref{Fig:4}. 
We also calculate radius $R_{MC}$  of the metastable cycle and its period $T_{MC}$, by performing both Monte-Carlo and 
SDE samplings. The radius scales proportionally to the fluctuation strength, $\sim \frac{1}{\sqrt{N}}$, and the period is near independent of $N$, 
as expected for a stochastic Hopf bifurcation \cite{hopf1}. 

The average lifetime of the metastable dynamics can defined, by using the largest eigenvalue $\lambda$ of the reduced
transition matrix $\tilde{\mathbf{Q}}^{1}$,  as $t_{\mathrm{life}} = 1/(1-\lambda)$ \cite{darroch}.
It  can be  then compared with the average extinction time \cite{meantime}, 
obtained by performing Monte-Carlo sampling of the Markovian dynamics.  The two times are in a perfect agreement;~see Fig.~\ref{Fig:5}.

\section{Periodically varying payoffs: a~metastable Floquet dynamics} 

\begin{figure}[t]
\includegraphics[width=0.47\textwidth]{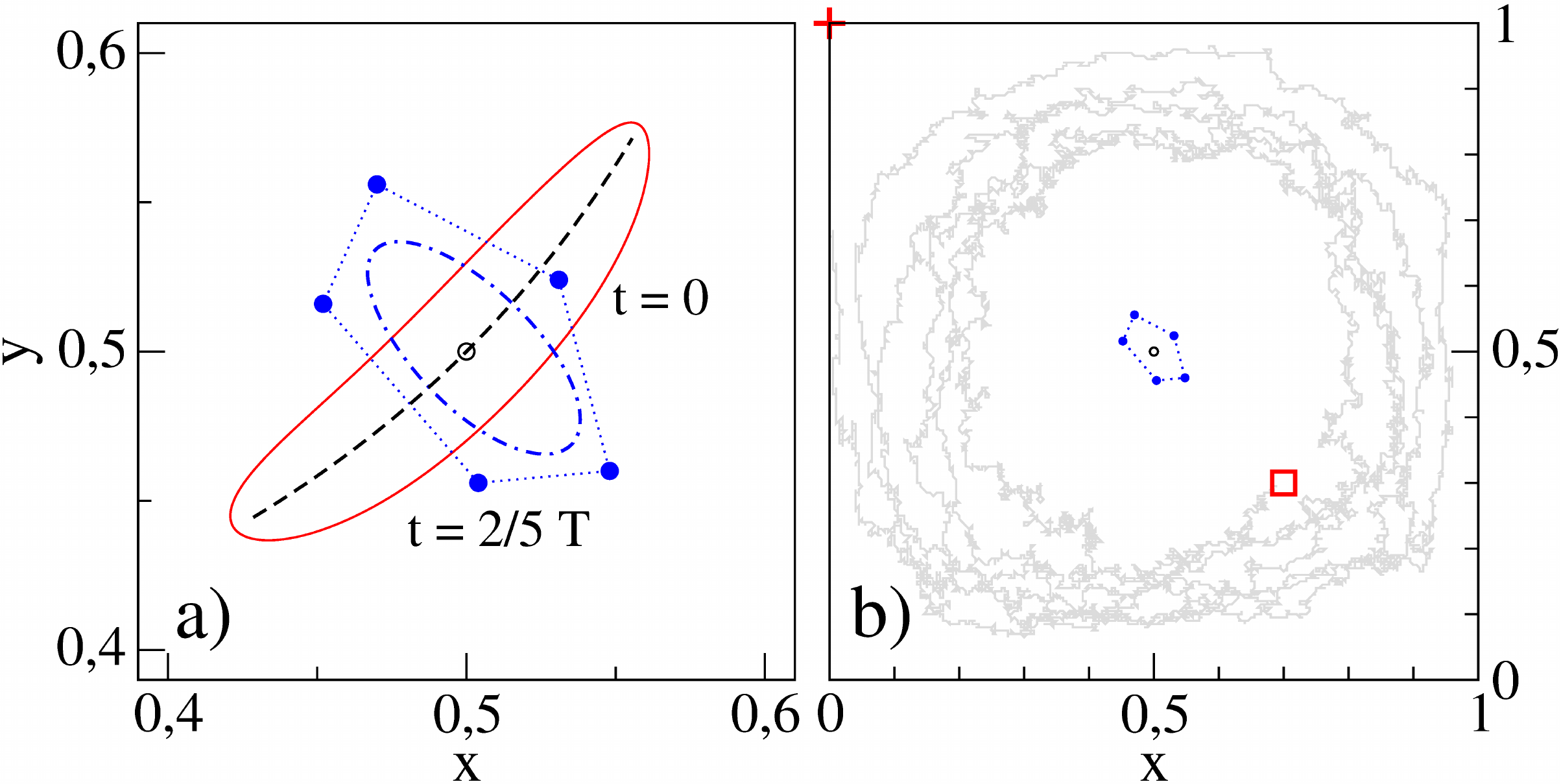}
\caption{(color online) \textbf{Evolution driven by a `Battle of Sexes' with periodically varying payoffs.} (a)
Period-one limit cycles of the mean-field dynamics, $N \rightarrow \infty$ with $\omega = 0.1$ (solid blue line) and $\omega = 0.01$ (solid red line), 
are localized near the point $\left(\frac{1}{2},\frac{1}{2}\right)$ (arrows indicate the direction of the motion). 
In the adiabatic limit, $\omega \rightarrow 0$, the mean-field cycle collapses to the set of the instantaneous Nash equlibria (dashed black line). 
Averages of the Floquet quasi-stationary distribution  $\mathbf{d}[t]$ for $N=200$, $\bar{x}(t)$ and $\bar{y}(t)$, evolve along a limit cycle (blues dots)
localized near the Nash equilibrium $\left(\frac{1}{2},\frac{1}{2}\right)$; (b) A stochastic trajectory (gray line) reveals 
the existence of a very de-localized transient dynamics. The trajectory was initiated at an interior  point (square) and ended up at an 
absorbing state (cross). The parameters are  $f = 0.5$, $\omega=0.1$, and
other parameters as in Fig.~\ref{Fig:2}.}\label{Fig:6}
\end{figure}

By introducing periodic variations into the payoffs of the `Battle of Sexes',
we find that the mean-field dynamics,~Eqs.~(\ref{Eq:replicatorx}-\ref{Eq:replicatory}), 
does not exhibit substantial changes even for a relatively large variation amplitude. For example, 
for a choice when $\epsilon(t) = \tilde{a}_{11}(t) = \tilde{b}_{22}(t)= f \cos(\omega t)$,
with $f = 0.5$ and $\omega = 2\pi/T$ and all other payoffs kept constant, 
we observe a period-one limit cycle localized near the Nash equilibrium of the constant payoff case, see Figs.~\ref{Fig:6}a). 
In the  limit $\omega \ll 1$, the cycle collapses to a set of  adiabatic Nash equlibria (dashed black lines in Fig.~\ref{Fig:6}a). Here an adiabatic Nash equilibrium is defined as the Nash equilibrium of a stationary game  with 'frozen' instantaneous values of payoffs, 
$\left\{x_{NE}(\epsilon) = \frac{2-\epsilon}{4 - \epsilon}, 
y_{NE}(\epsilon) = \frac{2}{4 + \epsilon}\right\}$.

\begin{figure}[t]
\includegraphics[width=0.45\textwidth]{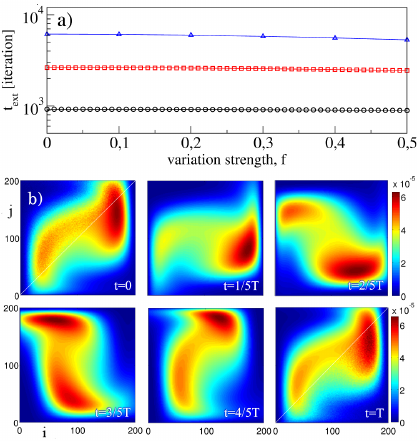}
\caption{(color online) \textbf{Metastable Floquet  dynamics.}
(a) The average extinction time $t_{\mathrm{ext}}$ as a function 
of the modulation strength $f$, for the population size $N = 50$ (dots), $100$ (squares), 
and $200$ (triangles). For every values of the variation strength$f$, $t_{\mathrm{ext}}$ was sampled with 
$10^5$ realizations per non-absorbing state; (b)~Stroboscopic snapshots of a quasi-stationary Floquet distributions $\mathbf{d}[t]$ for different time
instances, $t = m\cdot \Delta t \in [0,T]$.
Distributions are obtained by diagonalizing the reduced propagators $\tilde{\mathbf{Q}}^{m}$ for $N=200$. 
Other parameters are as in Fig.~\ref{Fig:2}.}\label{Fig:7}
\end{figure}

Figure~\ref{Fig:7}a shows the average extinction time $t_\mathrm{ext}$ as a function of  variation strength $f$ for $\omega = 0.1$.
In order to compare the evolution for different population size $N$, we keep $T$ constant and scale the time step as $\Delta t = \frac{2\pi}{N}$.
Another words, in terms of index $m$, the number of the rounds needed for one variation cycle is $M = \omega N$.

As it can be seen from Fig.~\ref{Fig:7}a, changes of $t_\mathrm{ext}$ with $f$ are not substantial for all three considered values of $N$. 
However,  the extinction times are  more than the order of magnitude shorter than in the case of constant payoffs (compare to Figure~\ref{Fig:5}). 
The Monte-Carlo sampling 
reveals the cycling transient dynamics which is much less localized around the the Nash equilibrium than the limit cycle corresponding to the mean-field
description; see Fig.~~\ref{Fig:6}b). Evidently, the dynamics of the means is not able to unreveal the whole complexity of the transient dynamics.
We are going now to generalize the idea of QDs to the case of periodically varying payoffs 
and demonstrate that the corresponding quasi-stationary state provides a deeper insight into the transient dynamics. 

The transition matrices, Eq.~(\ref{Eq:supermatrix}), 
are round-specific now and form a set $\{\tilde{\mathbf{S}}^{m}\}$, $m=1,\cdots,M$ (recall that after $M=T/\Delta t$ 
rounds the payoffs return to their initial values). The propagator over the interval $[0,t]$, $0 < t < T$, is 
the product $\tilde{\mathbf{U}}(0,t) = \prod_{k=1}^{m}\tilde{\mathbf{S}}^{k}$ with $m =t/\Delta t$. 
All the propagators, including the period-one propagator $\tilde{\mathbf{U}}(0,T) = \tilde{\mathbf{U}}_T$, 
have the same structure as the matrix in~Eq.~(\ref{Eq:supermatrix}). 

We introduce the quasi-stationary distribution of $\tilde{\mathbf{U}}_T$, $\mathbf{d}[0]$. 
It is a \textit{Floquet state} \cite{floquet,floquetH} of the reduced period-one propagator $\tilde{\mathbf{U}}_T^{r}$,
obtained by replacing,  in the definition of the propagator, transition matrix $\tilde{\mathbf{S}}^{m'}$ with reduced transition matrix $\tilde{\mathbf{Q}}^{m'}$. 
The maximal-eigenvalue  eigenvector of the matrix, after the reshuffling,  leads to probability distribution $\mathbf{d}[0] = \mathbf{d}[T]$. This 
is a stroboscopic snapshot of Floquet QD which periodically evolves, $\mathbf{d}[t] = \mathbf{d}[t+T]$, being locked by
payoff variations, see Fig.~\ref{Fig:7}a. For any value of $t$, $0 < t <T$, QD $\mathbf{d}[t]$ 
can be obtained by acting with the reduced propagator $\tilde{\mathbf{U}}^{r}(0,t)$ on the distribution $\mathbf{d}[0]$.

The structure and dynamics of the Floquet QD explains the dramatic shortening of the average extinction times. 
Namely, while in the constant limit, QD  $\mathbf{d}(x,y)$ was localized near 
the Nash equilibrium, i.e., at the maximal distance from the absorbing boundaries, both maxima of the Floquet QD 
skim very close to the boundary, see~Fig.~\ref{Fig:7}b. In the ecological context, this  can be considered as
a periodic sequence of population bottlenecks \cite{dawkins}. A better  articulated  biological interpretation of this effect is intriguing but  
goes outside the scope of our work.

Finally, we compare an average lifetime extracted from the Floquet QD with the result of the sampling of the average extinction time. 
The average lifetime of $\mathbf{d}[t]$ is defined by using the largest eigenvalue $\lambda_T$, $0<\lambda_T<1$, 
of  matrix $\tilde{\mathbf{U}}^{r}_T$. In order to compare it with the lifetime introduced for the case of constant payoffs, 
we define the mean single-round exponent as $\bar{\lambda}_T = \sqrt[T]{\lambda_T}$. The average 
lifetime is defined then as $t_{\mathrm{life}} = 1/(1-\bar{\lambda}_T)$. The obtained values  are in a perfect agreement (within the sampling error) 
with the values presented on~Fig.~\ref{Fig:7}a.

Different from the constant payoff case, the SDE approach, when realized numerically by using the first order Euler - Maruyama scheme,
works poorly in this case. In particular, the trajectory often crosses the boundary of the unit square thus making 
results meaningless \cite{note2}.
Even though the noise strength is strictly zero along the adsorbing boundary, it is not enough to prevent the trajectory from crossing the latter.
A numerical integration of  stochastic equations with multiplicative noise and time-dependent coefficients is a non-trivial task and its realization
demands more sophisticated higher-order schemes \cite{stoch}.

\section{Conclusions} 

By implementing the idea of quasi-stationary distributions \cite{darroch, QS}, we demonstrated that the transient (before the fixation) 
game-driven evolution of a large but finite population can be very different from the mean-field replicator picture. 
In the case of a game with constant payoffs, the transient evolution corresponds to a noisy cycling dynamics
which can be reproduced with a system of stochastic differential equations. Within this framework, the dynamics 
can be interpreted as a stochastic Hopf bifurcation \cite{arnold,hopf1,baxendale,kurths}

There are ongoing discussions on the role of  ``stochasticity in population cycling''~\cite{cycling0, cycling,cycling1} and the phenomenon of
``noise-sustained oscillations around otherwise stable equilibria'' was addressed recently in the ecological context \cite{cycling}.

It is important therefore to demarcate this type of cycling from the well-know 
phenomena  observed in many EGT models~\cite{hofbauer}, especially in system driven by games with so-called "cycling dominance"\cite{Szolnoki,szabo2016} (Rock–Paper–Scissors  is perhaps the most celebrated example). This phenomena
goes under different names, depending on the interpretation, e.g., "evolutionary cycling" \cite{c1}, "cycles of cooperation and defection" \cite{c2}, and "oscillating tragedy of the commons" \cite{c3}. On the formal level, the corresponding models demonstrate cycling behaviour in the thermodynamics limit, and the origin of the cycling  is a limit-cycle dynamics exhibited by the corresponding mean-field system. 

In contrast, in our models oscillations are absent in the mean-field picture due to the marginal stability of the fixed point. Our findings can be considered as an extension of the results by McKane and Newman~\cite{cycling0}. 
Namely, McKane and Newman highlighted the existence of a cycling behavior in a finite predator-prey model while the corresponding 
mean-field system is characterized, similar to adjusted replicator equations we considered, Eqs.~(\ref{Eq:replicatorx} - \ref{Eq:replicatory}), 
by a single marginally stable stable fixed point (an attracting point with zero Lyapunov exponents). Their main finding is that the mechanism responsible for the appearance of the cycling 
is internal and it is a ``demographic stochasticity inherent in discrete birth, death, and predation events'' \cite{cycling0}. 
Our consideration however new aspects, such as the essentially 
transient character of this cycling dynamics and finiteness of its lifetime.

By introducing periodically varying payoffs, we demonstrated the existence of a metastable, periodically-changing probability distribution, 
which cannot be deduced from the  evolution of the means. To be more specific,
the mean-field system yields the limit cycle strongly localized near the 
Nash equilibrium of the average (in terms of the paoffs) game; see Fig.~6b. In the case of finite-size dynamics with $N \gg 1$, it is usually expected that the corresponding probability distribution is localized on the cycle and the localization becomes stronger upon the increase of $N$. This is not so in the model with periodically varying payoffs: The corresponding  probability distribution is multi-modal and its first moments, $\bar{x}(t)$ and $\bar{y}(t)$, Eq.~(\ref{Eq:means}), do not reflect 
the complex shape and  dynamics of the distribution.
For example, during a single round of modulations,
maxima of the distribution pass very close to the absorbing border (see Fig.~7b) so that the probability of extinction for the corresponding player species is high. This, however, cannot be guessed by looking at the evolution of $\bar{x}(t)$ and $\bar{y}(t)$.

The idea that Floquet theory can be used to model the evolution of ecological systems subjected to time periodic environmental variations,
has recently been emphasized in Refs.~[\onlinecite{floquet_ecol,floquet_ecol1}]. However, it was used to  analyze  asymptotic 
regimes of models described with a set of deterministic linear differential equations, in the spirit of the conventional Floquet theory \cite{floquet}. 

Other potential applications of quasi-stationary Floquet distributions include modeling of periodically modulated finite systems with complex 
kinetics. For example, transient Floquet states can be related to a transient single-cell gene-expression dynamics, when the chemical kinetics of molecule 
ensembles is modulated by the inner-cell circadian rhythm \cite{cyrcadian, cyrcadian1}. 

Finally, we would like to mention a possibility of extension of the QS concept to quantum systems. Quantum channels \cite{qc}, also known as  "completely-positive trace-preserving maps" and "quantum operations" \cite{nielsen}, which can transform a quantum state (density operator) into another quantum state while preserving some important properties, are conventionally accepted as a quantum generalization  of Markov chains. Formalization of quasi-stationary quantum states and figuring out their 
relations to metastable \cite{les} and Floquet \cite{denisov} states of open quantum systems is a thought-provoking perspective.

\section*{Acknowledgment}
The authors acknowledge support of the Russian Foundation for Basic Research, grants No. 18-32-20221  and 20-32-90202,
and  and Lobachevsky University Center of Mathematics (M. I and O. V.). 
J. T acknowledges support by the Institute for Basic Science in Korea (IBS-R024-Y2). Numerical simulations were performed on 
the Lobachevsky supercomputer (Lobachevsky University, Nizhny Novgorod). 

\textbf{Data Availability}. The data that support the findings of this study are available from the corresponding author upon reasonable request.

\makeatletter \renewcommand{\fnum@figure}{{\bf{\figurename~S\thefigure}}}
\setcounter{figure}{0}
\setcounter{equation}{0}
\renewcommand{\theequation}{A\arabic{equation}}

\section*{Appendix: transition tensor}
Here we describe the transition fourth-order tensor $S^{m}(i,j,i',j')$ 
in terms of the rates [$T_{A}^{+,-}(i,j,t)$ and $T_{B}^{+,-}(i,j,t)$] 
for populations $A$ and $B$, given by Eq.~\eqref{Eq:rates} in the main text. 
The stochastic Moran process can be expressed as a Markov chain 
\begin{widetext}
\begin{eqnarray} 
p^{m+1}(i,j) &=& \left[1-T_{A}^{+}(i,j)-T_A^{-}(i,j)\right]\left[1-T_B^{+}(i,j)-T_B^{-}(i,j)\right]p^{m}(i,j)
+ T_B^{-}(i,j+1)\left[1-T_A^{-}(i,j+1)-T_A^{+}(i,j+1)\right]p^{m}(i,j+1) \nonumber \\
&& + T_B^{+}(i,j-1)\left[1-T_A^{-}(i,j-1)-T_A^{+}(i,j-1)\right]p^{m}(i,j-1)
+ T_A^{-}(i+1,j)\left[1-T_B^{-}(i+1,j)-T_B^{+}(i+1,j)\right]p^{m}(i+1,j) \nonumber \\
&& + T_A^{+}(i-1,j)\left[1-T_B^{-}(i-1,j)-T_B^{+}(i-1,j)\right]p^{m}(i-1,j)
+ T_A^{-}(i+1,j+1)T_B^{-}(i+1,j+1) p^{m}(i+1,j+1) \nonumber \\ 
&& + T_A^{+}(i-1,j+1)T_B^{-}(i-1,j+1) p^{m}(i-1,j+1)
+ T_A^{-}(i+1,j-1)T_B^{+}(i+1,j-1) p^{m}(i+1,j-1) \nonumber \\
&& + T_A^{+}(i-1,j-1)T_B^{+}(i-1,j-1) p^{m}(i-1,j-1). 
\end{eqnarray}
\end{widetext}

Above we have suppressed the time index $t=m\Delta t$ for all the rates.

The above equation can be recast into
\begin{eqnarray}
p^{m+1}(i,j) &=& \sum_{i',j'} S^{m}(i,j,i',j') p^{m}(i',j'), 
\end{eqnarray}
where the fourth-order tensor $S^{m}(i,j,i',j')$ is given by
\begin{widetext}
\begin{eqnarray}
S^{m}(i,j,i',j') &=& \left[1-T_A^{+}(i',j')-T_A^{-}(i',j')\right]\left[1-T_B^{+}(i',j')-T_B^{-}(i',j')\right]\delta_{i',i}\,\delta_{j',j}
+ T_B^{-}(i',j')\left[1-T_A^{-}(i',j')-T_A^{+}(i',j')\right]\delta_{i',i}\,\delta_{j',j+1} \nonumber \\
&& + T_B^{+}(i',j')\left[1-T_A^{-}(i',j')-T_A^{+}(i',j')\right]\delta_{i',i}\,\delta_{j',j-1}
+ T_A^{-}(i',j')\left[1-T_B^{-}(i',j')-T_B^{+}(i',j')\right]\delta_{i',i+1}\,\delta_{j',j} \nonumber \\
&& + T_A^{+}(i',j')\left[1-T_B^{-}(i',j')-T_B^{+}(i',j')\right]\delta_{i',i-1}\,\delta_{j',j}
+ T_A^{-}(i',j')T_B^{-}(i',j')\delta_{i',i+1}\,\delta_{j',j+1} \nonumber \\ 
&& + T_A^{+}(i',j')T_B^{-}(i',j')\delta_{i',i-1}\,\delta_{j',j+1}
+ T_A^{-}(i',j')T_B^{+}(i',j')\delta_{i',i+1}\,\delta_{j',j-1}
+ T_A^{+}(i',j')T_B^{+}(i',j')\delta_{i',i-1}\,\delta_{j',j-1},
\end{eqnarray}
\end{widetext}
with the indices $i,j,i',$ and $j'\in \{0, \cdots, N\}$. Using the bijection  $k=(N-1)j+i$ and $l=(N-1)j'+i'$, we obtain the required matrix form, Eq.~\eqref{Eq:supermatrix} in the main text.




\begin{thebibliography}{1000}
\bibitem{mutations} A mutation is a change of the strategy outside the adaptation processes.
\bibitem{note} For a population with $N$ players and $n$ possible strategies, the number of states is equal
to the number of different ways to distribute $N$ indistinguishable balls into $n$ distinguishable boxes, that is $C(N+n-1,n-1)$.
In the case of two antagonistic populations playing against each other, this number squares. In the simplest case $n=2$,
with two populations of the size $N=200$ each, we have to deal with a Markov chain with $200^2= 40 000$ states.
\bibitem{transition} We follow the convention that the stochastic matrix acts on the probability vector from the right.
\bibitem{note2} We were not able to avoid the crossing for $N=400$ 
even with the integration step $dt = 10^{-5}$.
\end{thebibliography}
\end{document}